\documentclass[prl,aps,twocolumn,showpacs,superscriptaddress,floatfix]{revtex4}
\pdfoutput=1

\usepackage[latin1]{inputenc} 
\usepackage{color}
\usepackage{graphicx,epsfig,epstopdf,amsmath}
\DeclareMathAlphabet{\mathpzc}{OT1}{pzc}{m}{it}
\begin{document} 
 
\title{Jahn-Teller versus quantum effects in the spin-orbital material LuVO$_3$} 

\author{M.~Skoulatos}
\affiliation{Physik-Department, Technische Universit\"at M\"unchen, D-85748 Garching, Germany}
\affiliation{Laboratory for Neutron Scattering and Imaging, Paul Scherrer Institute, CH--5232 Villigen, Switzerland}

\author{S.~Toth}
\affiliation{Laboratory for Neutron Scattering and Imaging, Paul Scherrer Institute, CH--5232 Villigen, Switzerland}

\author{B.~Roessli}
\affiliation{Laboratory for Neutron Scattering and Imaging, Paul Scherrer Institute, CH--5232 Villigen, Switzerland}

\author{M.~Enderle}
\affiliation{Institut Laue-Langevin - BP 156, F-38042 Grenoble C\'edex 9, France}

\author{K.~Habicht}
\affiliation{Helmholtz-Zentrum Berlin f\"ur Materialien und Energie, Lise-Meitner-Campus, D-14109 Berlin, Germany}

\author{D.~Sheptyakov}
\affiliation{Laboratory for Neutron Scattering and Imaging, Paul Scherrer Institute, CH--5232 Villigen, Switzerland}

\author{A.~Cervellino}
\affiliation{Swiss Light Source, Paul Scherrer Institute, CH--5232 Villigen, Switzerland}

\author{P.G.~Freeman}
\affiliation{Institut Laue-Langevin - BP 156, F-38042 Grenoble C\'edex 9, France}
\affiliation{Laboratory of Quantum Magnetism, Ecole Polytechnique F\'{e}d\'{e}rale de Lausanne (EPFL), CH-1015 Lausanne, Switzerland}

\author{M.~Reehuis}
\affiliation{Helmholtz-Zentrum Berlin f\"ur Materialien und Energie, Lise-Meitner-Campus, D-14109 Berlin, Germany}

\author{A.~Stunault}
\affiliation{Institut Laue-Langevin - BP 156, F-38042 Grenoble C\'edex 9, France}

\author{G.J.~McIntyre}
\affiliation{Institut Laue-Langevin - BP 156, F-38042 Grenoble C\'edex 9, France}

\author{L.D.~Tung}
\affiliation{Department of Physics, University of Liverpool, Crown Street, Liverpool, L69 7ZE, UK}

\author{C.~Marjerrison}
\affiliation{Laboratory for Developments and Methods, Paul Scherrer Institute, CH--5232 Villigen, Switzerland}

\author{E.~Pomjakushina}
\affiliation{Laboratory for Developments and Methods, Paul Scherrer Institute, CH--5232 Villigen, Switzerland}

\author{P.J.~Brown}
\affiliation{Institut Laue-Langevin - BP 156, F-38042 Grenoble C\'edex 9, France}

\author{D.I.~Khomskii}
\affiliation{II. Physikalisches Institut, Universit\"at zu K\"oln, Z\"ulpicher Str. 77, D--50937 K\"oln, Germany}

\author{Ch.~R\"uegg}
\affiliation{Laboratory for Neutron Scattering and Imaging, Paul Scherrer Institute, CH--5232 Villigen, Switzerland}
\affiliation{DPMC--MaNEP, University of Geneva, CH-1211 Geneva 4, Switzerland}

\author{A.~Kreyssig}
\affiliation{Ames Laboratory, US DOE, Iowa State University, Ames, Iowa 50011, USA}
\affiliation{Department of Physics and Astronomy, Iowa State University, Ames, Iowa 50011, USA}

\author{A.I.~Goldman}
\affiliation{Ames Laboratory, US DOE, Iowa State University, Ames, Iowa 50011, USA}
\affiliation{Department of Physics and Astronomy, Iowa State University, Ames, Iowa 50011, USA}

\author{J.P.~Goff}
\affiliation{Department of Physics, Royal Holloway, University of London - Egham, Surrey TW20 0EX, UK}

\date{\today} 

\begin{abstract} 

We report on combined neutron and resonant x-ray scattering results, identifying the nature of the spin-orbital ground state and magnetic excitations in LuVO$_3$ as driven by the orbital parameter. In particular, we distinguish between models based on orbital Peierls dimerization, taken as a signature of quantum effects in orbitals, and Jahn-Teller distortions, in favor of the latter. In order to solve this long-standing puzzle, polarized neutron beams were employed as a prerequisite in order to solve details of the magnetic structure, which allowed quantitative intensity-analysis of extended magnetic excitation data sets. The results of this detailed study enabled us to draw definite conclusions about classical vs quantum behavior of orbitals in this system and to discard the previous claims about quantum effects dominating the orbital physics of LuVO$_3$ and similar systems.

\end{abstract} 
 

\maketitle 
Geometrically frustrated magnetism with its crucial role of quantum effects in purely spin-systems is a well-established field with new, exotic phases emerging \cite{lacroix}. Can we, however, have a similar situation in a completely different context, namely, in orbital physics? A positive answer to this question would open up a new field and define a new \textit{class of materials}.
\\
\indent
The possible role of quantum fluctuations in orbital physics is a very interesting and important question. For small objects such as Jahn-Teller (JT)-active molecules or isolated JT impurities in solids, such quantum effects are well known and constitute a big field of vibronic effects in JT physics \cite{englman,bersuker}. On the other hand, in concentrated solids we practically always ignore these effects and treat orbitals (quasi)classically. Therefore all the more exciting were the suggestions \cite{khaliullin1p5,khaliullin2} that orbitals may behave as essentially quantum objects, in particular in some perovskite vanadates \cite{sirker, ulrich}, up to the formation of orbital singlets in YVO$_3$ \cite{ulrich}. If true, it would have opened a big new class of phenomena and novel group materials with quite nontrivial properties.
\\
\indent
However there are also arguments \cite{khomskii,khomskiiScripta}, that the situation with quantum effects in  orbitals may be not so simple and not exactly analogous to that in spin systems. It is predominantly connected with the intrinsically strong orbital--lattice coupling, as a result of which orbital degrees of freedom become "heavy", essentially classical (or one needs to treat also lattice vibrations quantum-mechanically, as is done in vibronic physics). Specifically, for YVO$_3$ Fang and Nagaosa \cite{fang} proposed an alternative, essentially classical explanation of the experimental findings of \cite{ulrich}, thus casting serious doubts on the importance of orbital quantum effects in $R$VO$_3$ -- practically the only real systems for which these effects were claimed to be observed.
\\
\indent
To clarify this important issue, we carried out a detailed experimental study of a material of the same group and with similar properties as YVO$_3$ -- LuVO$_3$, using polarized neutron and resonant x-ray scattering.
\begin{figure}[t!]
\centering
\includegraphics[scale=1]{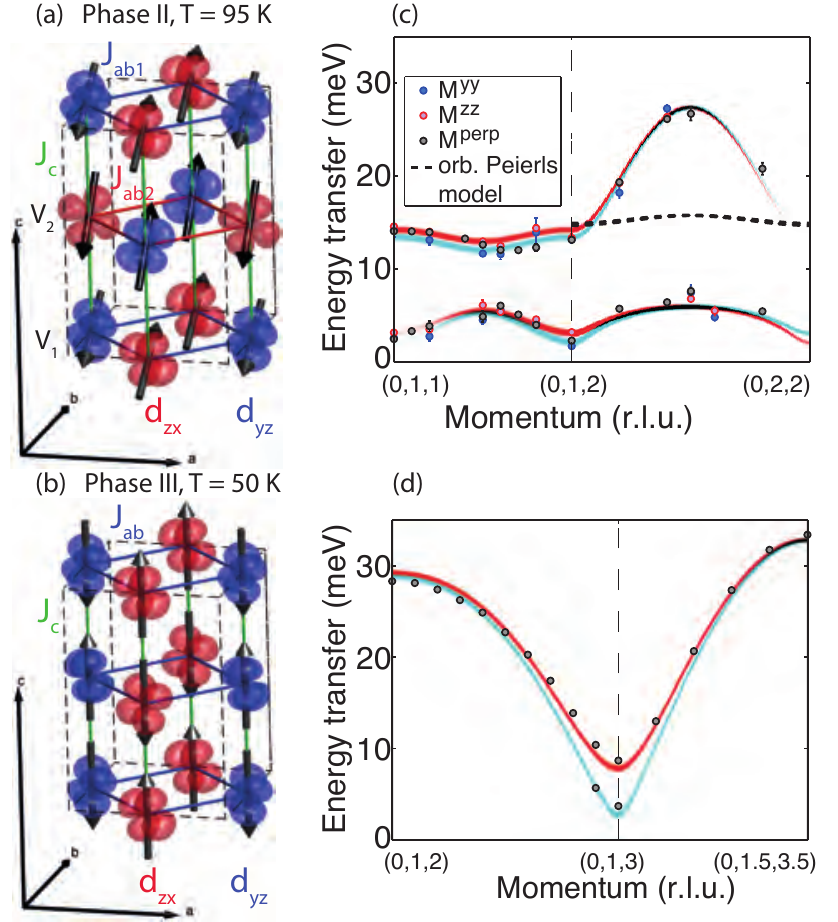}
\caption{Magnetic ground state configurations (a,b), spin waves (c,d) and relevant exchange coupling schemes for the two magnetic phases of LuVO$_3$. The magnetic (orbital) structures are derived from neutron polarimetry (resonant x-ray scattering) data while the spin waves were measured using neutron spectroscopy.
Calculations shown in (c) are for the Jahn-Teller model (solid lines with line width indicating the calculated spectral weight) and orbital-Peierls model (dashed line), as discussed in the text. Colors indicate the different polarisation states.}
\label{fig1}  
\end{figure}
\\
\indent
Specifically, the intriguing proposal made for the $R$VO$_3$ compounds was the existence of a novel state - an orbital-Peierls dimerization in the \textit{c}-direction \cite{sirker,ulrich}.
This proposal was based on the observation of a gap in the spin-wave spectrum along $L$, which implied an alternation of exchange interaction between consecutive layers in the $c$-direction.
The alternative proposal of \cite{fang} used the more conventional picture of JT distortion and orbital order (OO) alternating between consecutive $ab$-planes.
However, both these theories produce virtually identical spinwave dispersions along the $c$-direction (see Fig. 1c, left part where both theories coincide).
The spin-wave behaviour of the two competing models can be drastically different, but only for carefully selected reciprocal lattice directions.
Our results along the [0K2] direction (Fig. 1c, right part) conclusively favor a model based on Jahn-Teller distortions rather than the orbital Peierls dimerisation for $R$VO$_3$.
Another limiting experimental factor in order to distinguish between these two pictures is the fact that the material exhibits structural, magnetic and orbital transitions at the same temperature.
Therefore, spin-polarized neutron and resonant x-ray beams are essential in order to distinguish between magnetic, orbital and structural signals.
This is of utter importance since a prerequisite to answer the main important question is to firstly determine the precise magnetic structure. Based on that one can make a quantitative analysis on the magnetic-excitation intensities of the relevant phase.
\\
\indent
LuVO$_3$ is the end-member of the $R$VO$_3$ family and has the smallest ionic radius, giving rise to increased octahedral distortions.
We therefore expect it to be most sensitive to a possible orbital-Peierls dimerisation.
The V$^{3+}$ ion ($S=1$) is in an octahedral environment of O atoms (within the perovskite structure) with the two $3d$ electrons occupying $t_{2g}$--orbitals.
One electron occupies the $xy$ orbital, and the other -- one of the two doubly-degenerate $xz$ or $yz$ orbitals (or their linear superposition).
In LuVO$_3$ the interplay between spin and orbital physics is responsible for the rich phase diagram indicated by bulk measurements \cite{miyasaka,fujioka}, as for the well studied YVO$_3$ \cite{ren1,ren2,noguchi,blake1,blake2,reehuis1,ulrich,bizen}.
Upon cooling, LuVO$_3$ first enters an orbitally ordered phase at $T_{OO}=177$ K (phase I) followed by magnetic ordering at $T_{SO1}=105$ K (phase II). Below that, yet another orbital-magnetic phase transition takes place, at $T_{SO2}=82$ K (phase III) \cite{miyasaka}.
This information, together with the phase numbering used throughout this text, is given to the top of Fig. 2. 
\\
\indent
Experimental setup details can be found in the Supplementary Information.
The resonant x-ray scattering (RXS) experiment, being sensitive to anisotropic properties of the tensorial cross-section, yields the charge forbidden Bragg reflections arising from the OO and shown in Figs. 2(a), 2(b). Specifically, $G$-type orbital ordering is revealed in phases I+II, while $C$-type is found for phase III (schematic diagrams of the order in Fig. 2). Ab-initio calculations show that electric dipole transitions dominate the cross-section in this case. 
The neutron data shown in Figs. 2(c) and 2(d) complement the x-ray data; they show primarily $C$-type spin ordering for phase II (but with a small admixture of $G$-type) and $G$-type spin ordering only for phase III.
This combined neutron-RXS result is in agreement with the Goodenough-Kanamori rules, see \cite{khomskii3}.
\begin{figure}[t]
\includegraphics[width=\linewidth]{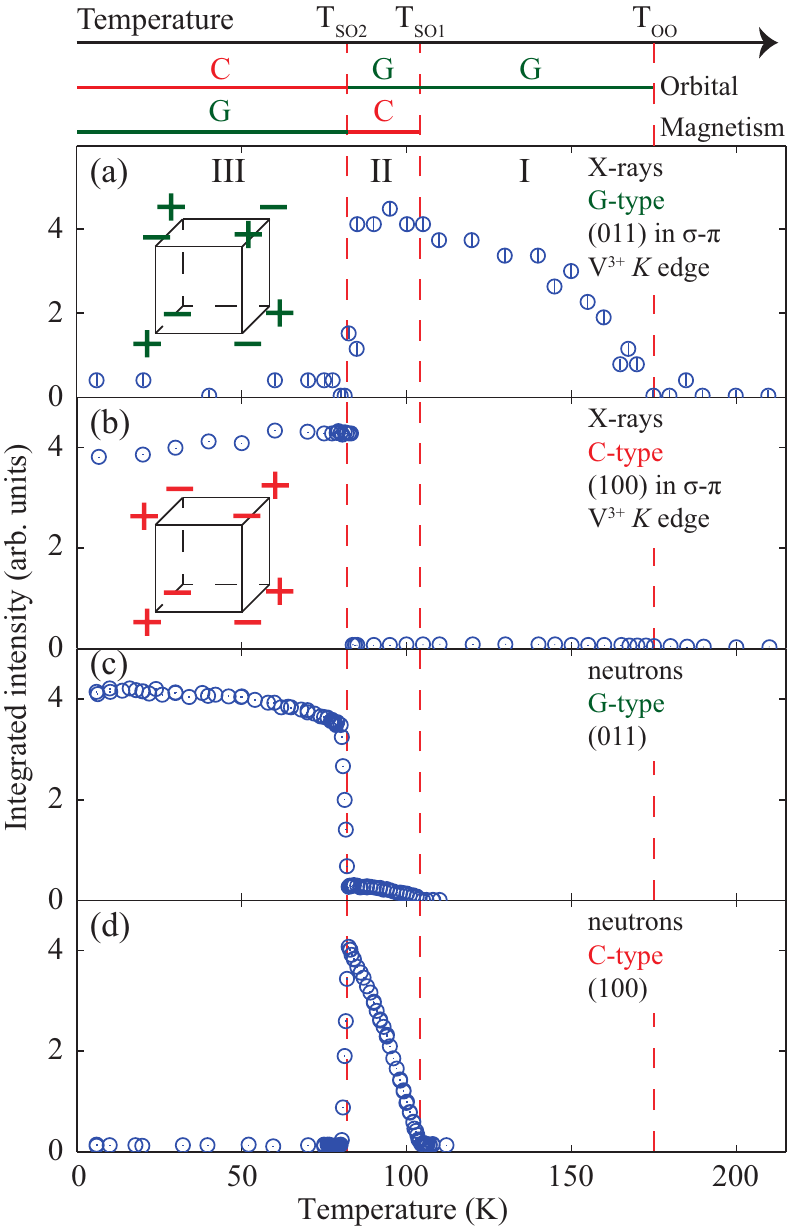}
\caption{Spin and orbital order parameters of LuVO$_3$ as a function of temperature, revealed by (a,b) x-ray and (c,d) neutron scattering. The complementarity of the techniques gives a clear picture of the ordered structures: (a) $G$-type orbital order for phases I, II and (b) $C$-type orbital order for phase III. At the same time, we observe (c) $G$-type spin order for phase III while (d) $C$-type spin order for phase II, in accordance with the Goodenough-Kanamori rules \cite{kugel}. Note the small $G$-type magnetic component in phase II, which gives an overall canted structure. An overview is schematically drawn on top of the figure. "+" and "-" refer to spins "up" and "down" or an orbital configuration, for instance "$d_{yz}$" and "$d_{zx}$" respectively.} 
\label{fig2} 
\end{figure}
\\
\indent
These measurements demonstrate that LuVO$_3$ is an antiferromagnet (AF) with \textbf{\textit{k}}=0. This, combined with structural phase transitions occurring at the magnetic ordering temperatures, necessitates the use of polarized neutrons to determine the magnetic structures in detail.
Spherical neutron polarimetry is used to measure polarization matrices for selected reflections in both magnetic phases, as discussed in the supplementary information.
The unpolarized intensity data (E5) were used for cross-check as well as for normalising the moment size.
\\
\indent
Phase III has a collinear G-type magnetic structure as plotted in Fig. 1(b), consistent with the $Pbnm$ orthorhombic space group. The magnetic moments are pointing purely along the crystallographic $C$-axis. For phase II, beam depolarisation arising due to orientation domains implies that the space group can no longer be $Pbnm$.
Furthermore, in order to have $G$-type orbital ordering, as observed from our resonant x-ray data, it is necessary to lose the mirror plane perpendicular to the $c$-axis, also inconsistent with $Pbnm$.
We could thus fit our data with the lower monoclinic $P2_1/b$ space group and a canted magnetic structure, as shown in Fig 1(a).
\\
\indent
In order to gain deeper insight into structural details affecting the precise environment of the V$^{3+}$ ions, high-resolution powder x-ray diffraction experiments were performed (see suppl. material for details).
The orbitals plotted in Fig. 1 follow precisely the octahedral tilts and distortions as determined from refinements of these high-resolution x-ray data.
\\
\indent
The spin wave dispersion was measured in both phases to help modelling the microscopic Hamiltonian.
Polarized neutrons were used to measure the dispersion in phase II (shown in Fig. 1c), in order to disentangle the spin waves from phonon modes, which show similar dispersions.
Further, we were able to separate magnetic excitations with different polarization states (M$^{yy}$ and M$^{zz}$, standard Blume-Maleev coordinate system notation \cite{blume}), which is important in the subsequent analysis.
Magnons in phase III were measured by unpolarized neutron spectroscopy.
\begin{table}[b]
\caption{Exchange parameters for both magnetic phases of LuVO$_3$. Positive sign corresponds to AF coupling.}
\begin{center}
\scalebox{1}{
\begin{tabular}{| p{4.4cm}  | p{3.2cm} | }
\hline
Phase III (meV) & Phase II (meV)\\
\hline
$J_{ab}=4.24(21)$   & $J_{ab1}=0.82(3)$ \\
                    & $J_{ab2}=5.99(3)$ \\
$J_{c}=5.95(19)$    & $J_{c}=-1.29(2)$  \\
$K=(-0.48(12),~-0.06(2),~0)$ & $K_1=(0,~0.66(5),~0)$  \\
                    & $K_2=(0,~0,~0.66(5))$  \\
\hline
\end{tabular}
}
\end{center}
\end{table}
\\
\indent
The spin  wave dispersion was modelled using linear spin-wave theory (LSWT) and the $SpinW$ library \cite{sandor} with a simple Hamiltonian of the form
\mbox{$H=\underset{<i,j>}{\Sigma}J_{ij}\textbf{S}_i \textbf{S}_j+H_{an}$}, where $H_{an}$ is the usual easy-axis single-site anisotropy term like $-KS_z^2$, but with the local easy axes different for different sites, see below.
$J_{ij}$ exchange parameters are shown in Fig. 1 (positive $J_{ij}$ denotes AF coupling).
Phase III is fitted with two Heisenberg exchange parameters and a single-ion anisotropy term (parameters given in Table 1), yielding a perfect agreement with our data, as shown in Fig. 1(d).
The single ion anisotropy term gives rise to two split modes as well as the overall gap of the system, found to be $E_{gap}=3.7(1)$ meV. The situation is more complex and very different, for the intermediate temperature phase II. Magnons are observed at both $Q=(0~1~2)$ and $Q=(0~1~1)$ due to the canted structure ($C+G$ types). This new phase is gapped, with four modes and a splitting  of $\approx$5 meV between the two sets of branches. The $c$ axis energy scale is reduced by more than a factor of two while the doubling of the number of modes is in accordance with two inequivalent V sites (lowering of the space group).
\\
\indent
In order to determine the Hamiltonian, in this more complex case, it is necessary to model data along a further direction ($b$ axis, Fig. 1(c)).
The dispersion along the $b$ axis rises to 27 meV, an energy which can only be explained by assuming the interaction scheme of Fig. 1(a) with two distinct $ab$ plane couplings alternating along $c$, as well as a ferromagnetic coupling along the $c$ axis \cite{fang}.
The parameters of this Hamiltonian which fit the data very precisely are also given in Table 1.
In order to account for the canting of the moments, two individual directions were assumed for the single ion anisotropy term: easy $b$ and $c$ axes for V$_1$ and V$_2$ respectively, located in alternating $ab$-planes in accordance with the $P2_1/b$ space group symmetry and the interaction scheme. Note that this model accounts not only for the measured spin wave energies and intensities, but also correctly indexes the M$^{yy}$/M$^{zz}$ polarization states and further respects the magnetic structure as well as the symmetry of the $P2_1/b$ space group.
An alternative orbital Peierls model with alternating exchange along the $c$-direction and a single coupling in the $ab$ plane would require a lowering of the space group due to loss of a mirror plane along the crystallographic $c$ axis. This model \cite{ulrich} was fitted to the energies of our dataset, but predicts an almost flat mode at 15 meV for the dispersion along the $b$ axis (see black dashed line in Fig. 1c) with zero intensity, contrary to our observation.
The direct comparison of both models with the data (Fig. 3) shows that indeed the coupling scheme of Fig. 1(a) (continuous red lines) is in excellent agreement with our data. In contrast, the orbital-Peierls model with exchange couplings alternating along $c$ (dashed black lines) shows large overall discrepancies in the spectral weight distributions.
The figure shows the calculated intensities after fitting the dispersion energies (Fig. 1), with a single scaling factor.
Note that the symmetry of the monoclinic $P2_1/b$ space group is consistent with  inequivalence of the bonds  in alternate $ab$ planes (3D JT model), but incompatible with inequivalence of the $c$ axis bonds between these planes (orbital-Peierls scenario).
This is in addition a very strong symmetry argument in favour of the former model.
\begin{figure}[t]
\includegraphics[width=\linewidth]{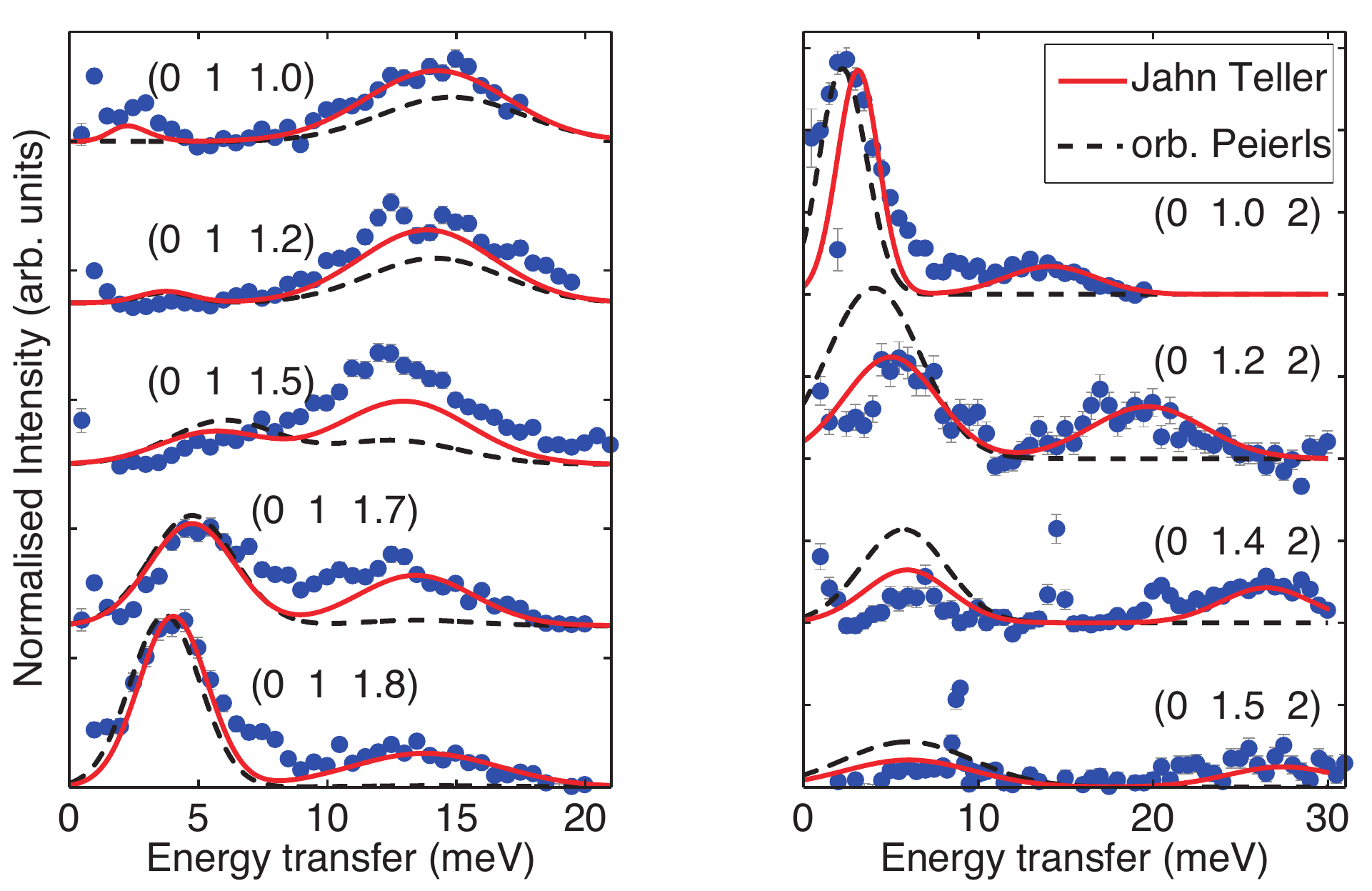}
\caption{Magnetic scattering intensity as measured in the intermediate phase of LuVO$_3$ ($T$=95 K). The directions shown are the $c^*$ and $b^*$-axes (left and right panels respectively). The exchange models being compared are either the 3D Jahn-Teller model (solid red lines) or the orbital Peierls model (dotted black lines).} 
\label{fig4}  
\end{figure} 
\\
\indent
Comparing phase III of LuVO$_3$ at low temperature with YVO$_3$ we find a direct agreement between the ground state and model Hamiltonian as discussed in Ref. \cite{ulrich}. The moment direction, inelastic gap-size and band-width match well. The main difference is a 30\% smaller ordered moment and variations of up to 40\% in the exchange couplings and anisotropy parameters 
which can be attributed to the details of lattice distortions and exchange pathways in each case.
However, the experimentally determined magnetic structure of phase II at intermediate temperature differs from all other $R$VO$_3$ members \cite{munoz,reehuis1,ulrich,reehuis2}. 
The use of polarized neutron diffraction enables precise determination of the moment direction ($bc$-plane), due to the $k=0$ magnetic propagation vector combined with structural distortions across the magnetic phase transitions.
Furthermore, by using full polarization analysis it is possible to predict the subtle lowering of the space group from orthorhombic to monoclinic, based on symmetry arguments and observation of orientation domains.
\\
\indent
By employing polarized neutron spectroscopy and, in particular, by making extended measurements in two inequivalent directions in reciprocal space, we were able to distinguish between model Hamiltonians in which either the exchange $J_{ab}$ or  $J_c$ alternate along the c-axis between two values (see Figures 1a, c and 3).
The clear evidence for the first model ($J_{ab}$ alternation) is the superiority of dispersion and intensity fits to this model, compared to the other ($J_c$ alternation). On top, this is in excellent agreement with $ab-initio$ theoretical calculations given in Ref. \cite{fang}.
The alternate $J_{ab}$ exchange parameters calculated for YVO$_3$ are 0.8 and 5.3 meV, and our fitted results for LuVO$_3$ are 0.8 and 6 meV, strongly supporting our conclusion.
These first-principle calculations \cite{fang} are based on the precise experimentally determined JT distortions of YVO$_3$ and related vanadates \cite{bordet,blake1,blake2,noguchi} which do indeed alternate between adjacent $ab$-planes, exactly in the same fashion as $J_{ab}$ in our Hamiltonian.
The $J_{ab}$ exchange parameter is very sensitive to these JT distortions and orbital ordering, because it depends on subtle competition between various exchange processes \cite{fang}.
This explains the large difference between the two alternating values of $J_{ab}$.
\\
\indent
Based on these new findings, we conclude that the orbital fluctuations --which are inherent to these systems-- are in fact suppressed by the JT distortions. In LuVO$_3$ this results in an overall 3D spin-orbital structure, rather than a quasi--1D orbital dimerized chain.
\\
\indent
Summarizing, we have carried out a detailed analysis of the interplay amongst spin, orbital and lattice degrees of freedom in the Mott insulator LuVO$_3$. By combining a variety of experimental methods, we are able to uniquely determine the magnetic and orbital states and to model the spin Hamiltonian in the two magnetic phases.
These results show that the features, attributed previously to  an orbital-Peierls state ("orbital-singlet", similar to spin-singlet dimers) which could have appeared due to quantum effects in orbitals, are in fact a consequence of the static orbital ordering and corresponding Jahn-Teller distortion.
Yet, we cannot rule out that the orbital quantum fluctuations may still be present in some form, maybe in different materials.
This question deserves further study.
\\
\\
ACKNOWLEDGEMENTS
\par
We are grateful to G. Khaliullin, O. Zaharko, Ch. Pfleiderer and S. Ward for invaluable discussions. We thank the sample environment teams at PSI, ILL and HZB, where these measurements were performed, for their expert assistance.
The research leading to these results has received funding from the European Community's Seventh Framework Programme (FP7/2007-2013) under grant agreement n. 290605 (PSI-FELLOW/COFUND).
The work of D. I. Khomskii was supported by the German program FOR 1346 and by the Cologne University via German Excellence Initiative.
Work at Ames Laboratory was supported by the US Department of Energy (DOE), Office of Science, Basic Energy Sciences, Materials Science and Engineering Division under Contract No. DE-AC02-07CH11358.
Use of the Advanced Photon Source, an Office of Science User Facility operated for the US DOE by Argonne National Laboratory under Contract No. DE-AC02-06CH11357.
XRPD data were collected at the X04SA-MS beamline of the SLS synchrotron at PSI.

\end{document}


\title{Supplementary Information for "Jahn-Teller versus quantum effects in the spin-orbital material LuVO$_3$"} 

\author{M.~Skoulatos}
\affiliation{Physik-Department, Technische Universit\"at M\"unchen, D-85748 Garching, Germany}
\affiliation{Laboratory for Neutron Scattering and Imaging, Paul Scherrer Institute, CH--5232 Villigen, Switzerland}

\author{S.~Toth}
\affiliation{Laboratory for Neutron Scattering and Imaging, Paul Scherrer Institute, CH--5232 Villigen, Switzerland}

\author{B.~Roessli}
\affiliation{Laboratory for Neutron Scattering and Imaging, Paul Scherrer Institute, CH--5232 Villigen, Switzerland}

\author{M.~Enderle}
\affiliation{Institut Laue-Langevin - BP 156, 38042 Grenoble C\'edex 9, France}

\author{K.~Habicht}
\affiliation{Helmholtz-Zentrum Berlin f\"ur Materialien und Energie, Lise Meitner Campus, D-14109 Berlin, Germany}

\author{D.~Sheptyakov}
\affiliation{Laboratory for Neutron Scattering and Imaging, Paul Scherrer Institute, CH--5232 Villigen, Switzerland}

\author{A.~Cervellino}
\affiliation{Swiss Light Source, Paul Scherrer Institute, CH--5232 Villigen, Switzerland}

\author{P.G.~Freeman}
\affiliation{Institut Laue-Langevin - BP 156, 38042 Grenoble C\'edex 9, France}
\affiliation{Laboratory of Quantum Magnetism, Ecole Polytechnique F\'{e}d\'{e}rale de Lausanne (EPFL), CH-1015 Lausanne, Switzerland}

\author{M.~Reehuis}
\affiliation{Helmholtz-Zentrum Berlin f\"ur Materialien und Energie, Lise Meitner Campus, D-14109 Berlin, Germany}

\author{A.~Stunault}
\affiliation{Institut Laue-Langevin - BP 156, 38042 Grenoble C\'edex 9, France}

\author{G.J.~McIntyre}
\affiliation{Institut Laue-Langevin - BP 156, 38042 Grenoble C\'edex 9, France}

\author{L.D.~Tung}
\affiliation{Department of Physics, University of Liverpool, Crown Street, Liverpool, L69 7ZE, UK}

\author{C.~Marjerrison}
\affiliation{Laboratory for Developments and Methods, Paul Scherrer Institute, CH--5232 Villigen, Switzerland}

\author{E.~Pomjakushina}
\affiliation{Laboratory for Developments and Methods, Paul Scherrer Institute, CH--5232 Villigen, Switzerland}

\author{P.J.~Brown}
\affiliation{Institut Laue-Langevin - BP 156, 38042 Grenoble C\'edex 9, France}

\author{D.I.~Khomskii}
\affiliation{II. Physikalisches Institut, Universit\"at zu K\"oln, Z\"ulpicher Str. 77, D--50937 K\"oln, Germany}

\author{Ch.~R\"uegg}
\affiliation{Laboratory for Neutron Scattering and Imaging, Paul Scherrer Institute, CH--5232 Villigen, Switzerland}
\affiliation{DPMC--MaNEP, University of Geneva, CH-1211 Geneva 4, Switzerland}

\author{A.~Kreyssig}
\affiliation{Ames Laboratory, US DOE, Iowa State University, Ames, Iowa 50011, USA}
\affiliation{Department of Physics and Astronomy, Iowa State University, Ames, Iowa 50011, USA}

\author{A.I.~Goldman}
\affiliation{Ames Laboratory, US DOE, Iowa State University, Ames, Iowa 50011, USA}
\affiliation{Department of Physics and Astronomy, Iowa State University, Ames, Iowa 50011, USA}

\author{J.P.~Goff}
\affiliation{Department of Physics, Royal Holloway, University of London - Egham, Surrey TW20 0EX, UK}

\date{\today} 

\maketitle

\indent
The measurements were performed on both powder samples and single-crystal fragments using synchrotron x-rays to determine the crystal structure and orbital order pattern. The instruments used were the X04SA-MS beamline at SLS, Villigen \cite{willmott} the 6ID-B resonant x-ray beamline, APS, Argonne.
The diffraction part was performed using $\lambda=0.620$ \AA. For the resonant x-ray experiment the vanadium $K$-edge ($E=5.482$ keV) was used in order to acess the orbital order wavevector transfer $Q$ required.
In order to probe the magnetic order parameter and excitations directly we have used diffraction and spectroscopy neutron scattering techniques respectively.
The four-circle diffractometer E5 at HZB, Berlin and the D3 diffractometer at ILL with the spherical neutron polarimetry option (CRYOPAD) were operated at $\lambda=2.38$ \AA~and $\lambda=0.825$ \AA~with a $^3$He spin analyser respectively.
The FLEX triple-axis spectrometer (TAS) at HZB \cite{skoul1,le1} was operated with a fixed final wavevector $k_f=1.55$ \AA$^{-1}$, and a Be-filter to reduce second-order contamination.
The thermal-neutron TAS IN8 and IN20 at ILL, Grenoble (the last with $XYZ$-polarization analysis) were used with fixed final wavevector $k_f=2.662$ \AA$^{-1}$ and a pyrolytic graphite filter for second-order contamination.
The large single crystal used in the spectroscopic measurements weighs 4 g.
\begin{table}[!h]
\caption{Representative measured and calculated polarization matrices used to determine the magnetic structures for both magnetic phases of LuVO$_3$.}
\begin{center}
\scalebox{0.8}{
\begin{tabular}{ccc|ccc|ccc|ccc}
\multicolumn{12}{l}{\textbf{phase III (low T phase)}} \\
\hline
\textit{h} & \textit{k} & \textit{l} & \multicolumn{3}{|c|}{initial P} & \multicolumn{3}{|c|}{Observed} & \multicolumn{3}{|c}{Calculated}\\
\hline
1 & 0 & 1 & 1 & 0 & 0  & -0.65(2)   & 0.07(1)   &  0.05(1) & -0.62 & 0.03 &  0.00 \\
1 & 0 & 1 & 0 & 1 & 0  & -0.00(1)   & 0.98(2)   & -0.02(1) &  0.00 & 1.00 &  0.00 \\
1 & 0 & 1 & 0 & 0 & 1  & -0.03(1)   & 0.05(1)   & -0.62(2) &  0.00 & 0.03 & -0.62 \\
\hline
3 & 0 & 1 & 1 & 0 & 0  & 0.68(6)   & 0(1)       & 0(1)     & 0.59  & -0.03 & 0.00 \\
3 & 0 & 1 & 0 & 1 & 0  & 0(1)      & 0.98(8)    & 0(1)     & 0.00  & 1.00  & 0.00 \\
3 & 0 & 1 & 0 & 0 & 1  & 0(1)      & 0(1)       & 0.79(7)  & 0.00  & -0.03 & 0.59 \\
\hline
1 & 0 & 3 & 1 & 0 & 0  & 0.99(1)   & 0(1)       & 0(1)     & 0.99  & 0.00 & 0.00 \\
1 & 0 & 3 & 0 & 1 & 0  & 0(1)      & 1.01(1)    & 0(1)     & 0.00  & 1.00 & 0.00 \\
1 & 0 & 3 & 0 & 0 & 1  & 0(1)      & 0(1)       & 0.99(1)  & 0.00  & 0.00 & 0.99 \\
\hline
1 & 0 & 1 & -1 & 0 & 0 & -0.63(2)  & -0.04(1)   &  0.02(1) & -0.62 & 0.03 &  0.00 \\
1 & 0 & 1 & 0 & -1 & 0 &  0.01(1)  &  0.97(2)   & -0.00(1) &  0.00 & 1.00 &  0.00 \\
1 & 0 & 1 & 0 & 0 & -1 & -0.03(1)  &  0(1)      & -0.62(2) &  0.00 & 0.03 & -0.62 \\
\hline
\multicolumn{12}{l}{\textbf{phase II (intermediate T phase)}} \\
\hline
H & K & L & \multicolumn{3}{|c|}{initial P} & \multicolumn{3}{|c|}{Observed} & \multicolumn{3}{|c}{Calculated}\\
\hline
0 & 1 & 0 & 1 & 0 & 0  & -1.00(10) &  0.05(7)   &  0.01(7)  & -1.00 & 0.00 & 0.00 \\
0 & 1 & 0 & 0 & 1 & 0  &  0.01(7)  &  1.02(10)  & -0.01(7)  &  0.00 & 1.00 & 0.00 \\
0 & 1 & 0 & 0 & 0 & 1  & -0.02(7)  & -0.01(7)   & -1.03(11) &  0.00 & 0.00 &-1.00 \\
\hline
1 & 0 & 0 & 1 & 0 & 0  & -1.00(7)  &  0.01(5)   &  0.02(5)  & -1.00 & 0.00 & 0.00 \\
1 & 0 & 0 & 0 & 1 & 0  & -0.00(4)  &  0.11(4)   &  0.06(4)  &  0.00 & 0.10 & 0.05 \\
1 & 0 & 0 & 0 & 0 & 1  & -0.03(4)  &  0.05(4)   & -0.10(4)  &  0.00 & 0.05 &-0.10 \\
\hline
1 & 0 &-2 & 1 & 0 & 0  & -1.00(17) &  0(1)      &  0(1)     & -1.00 & 0.00 & 0.00 \\
1 & 0 &-2 & 0 & 1 & 0  &  0(1)     & -0.42(13)  &  0(1)     &  0.00 &-0.41 & 0.05 \\
1 & 0 &-2 & 0 & 0 & 1  &  0(1)     &  0(1)      &  0.38(12) &  0.00 & 0.05 & 0.41 \\
\hline
1 & 0 & 2 & 1 & 0 & 0  & -1.00(15) &  0(1)      &  0(1)     & -1.00 & 0.00 & 0.00 \\
1 & 0 & 2 & 0 & 1 & 0  &  0(1)     & -0.44(11)  &  0.11(9)  &  0.00 &-0.41 & 0.05 \\
1 & 0 & 2 & 0 & 0 & 1  &  0(1)     &  0.13(10)  &  0.37(11) &  0.00 & 0.05 & 0.41 \\
\hline
\end{tabular}
}
\end{center}
\end{table}
\\
\indent
The magnetic structure of LuVO$_3$ was determined in both phases II and III by using spherical neutron polarimetry. The resulting polarisation matrices were modelled with the code $Mufit$ \cite{bertrand} based on the Blume-Maleev equations \cite{blume,maleev}. Each element of the matrix was averaged over the different domains, the populations of which were treated as free parameters and fitted by the code. Table SI gives the data and the fits for a selection of reflections.
\\
\indent
Phase III is a collinear $G-$type AF with moments pointing along the crystallographic $c-$axis as shown in Fig. 1(b) of the main text. This is consistent with the $Pbnm$ orthorhombic space group. The  moment  vector on each V site is $M_G^{III}=[0~0~1.20(2)]~\mu _B$. The relative populations of the 180$^{\circ}$ domains that break time-reversal symmetry is 48(3)\%-52(3)\%.
Phase II is more complex, partly because it combines $C-$ and $G-$type ordering.
We find $M_C^{II}=[0~0.66(5)~0.74(5)]~\mu _B$ and $M_G^{II}=[0~0.32(8)~\pm 0.36(8)]~\mu _B$ for the two contributions, i.e. the moments are canted and lie in the $bc-$plane. The '$\pm$' ambiguity arises from the existence of equally populated orientation domains, making it impossible to correlate the orientation domains of the $G$ and $C$ modes.
However, the '+' solution results in a moment size larger than  possible for a V$^{3+}$ free ion. As shown in Fig. 1(a), the '--' solution gives a lattice with a uniform overall moment size per site $M_{total}^{II}=1.10~\mu _B$.
Neither the spin orientation derived from the polarized neutron data nor the combination of $C$ and $G$ modes is compatible with space group $Pbnm$, but rather point to a lowering of the crystal symmetry to the monoclinic space group $P2_1/b$. This is consistent with the presence of orientation domains (fitted populations 47(3)\%-53(3)\%).
The neutron polarimetry data and $P2_1/b$ space group of this phase are consistent with the symmetry of the Hamiltonian modelling our spin-wave data, as discussed in the main text.
\begin{figure}[t]
\includegraphics[width=\linewidth]{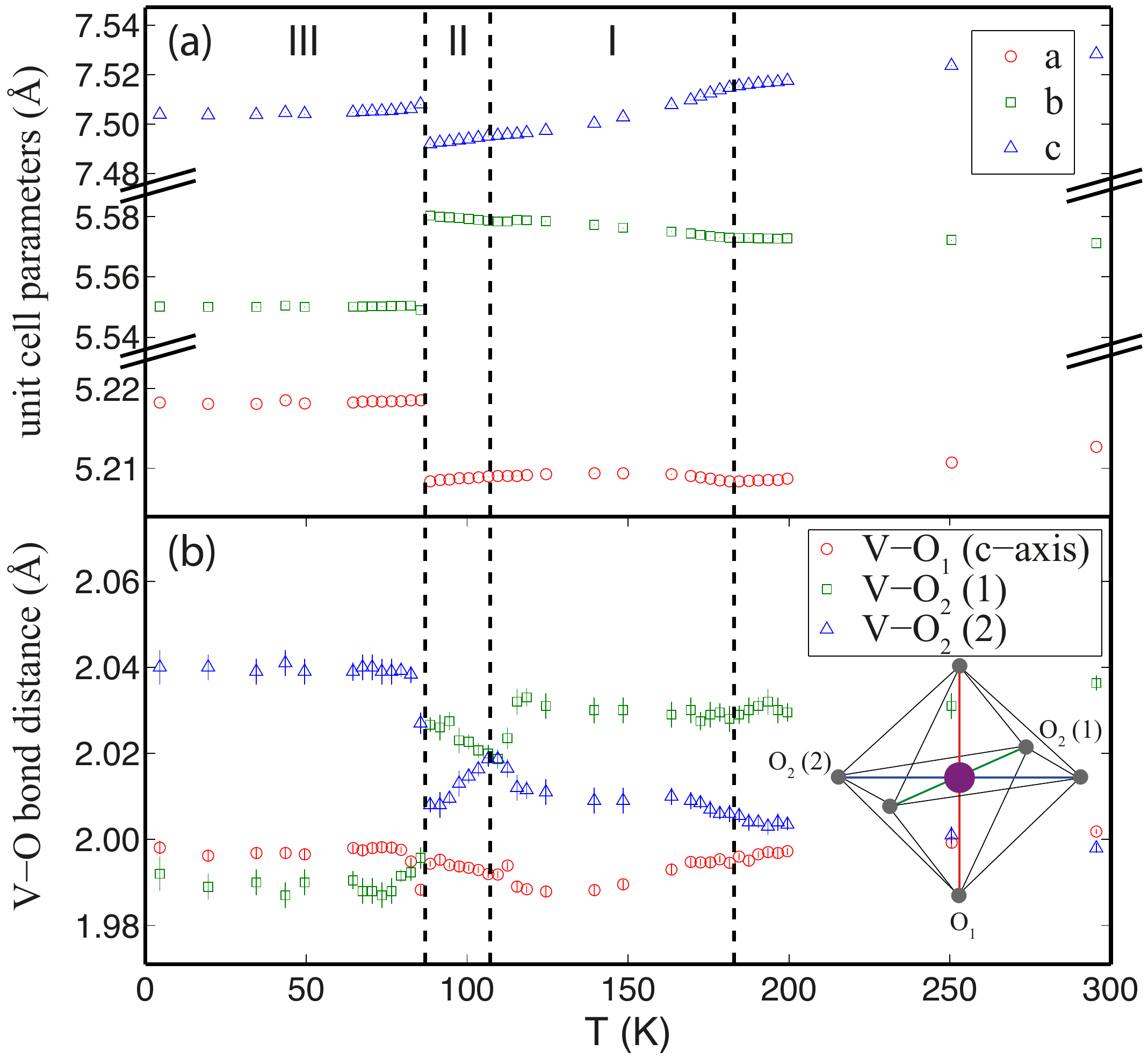}
\caption{Variation in the octahedral environment of V$^{3+}$ in LuVO$_3$ as a function of temperature: (a) lattice parameters and (b) V-O bond lengths refined from high-resolution synchrotron x-ray data.}
\label{fig3}  
\end{figure}
\indent
X-ray powder diffraction patterns were collected at 60 different temperatures and fitted using the FullProf program \cite{Rodriguez-Carvajal}.
The phase transitions are visible in the unit cell parameters plotted in Fig. SI(a). 
It is most instructive, however, to inspect the V-O distances shown in Fig. SI(b), which are those dominating the super-exchange interactions. The V-O1 (apical) distance stays relatively constant as a function of temperature, suggesting that the occupancy of the $d_{xy}$ orbital is unchanged. By contrast the lengths of the two distinct V-O2 bonds change quite drastically across the III$\rightarrow$II phase transition and the difference between them changes sign pointing to a corresponding swap in the occupancies of the $d_{xz}$ and $d_{yz}$ states. Their absolute difference is minimised in phase II  and changes sign at $T=100$ K. We can therefore expect drastic changes in the effective Hamiltonian across the transition and orbital fluctuations in phase II, as predicted by theory \cite{khaliullin1}.
Structural refinements of our high-resolution x-ray data with the two space groups ($Pbnm$ or $P2_1/b$) give indistinguishable fit quality and $\chi^2$. In other words, x-ray diffraction alone is unable in this particular case to solve the crystallographic structure, $even$ with the highest possible resolution available since the monoclinic splitting is too small. Our case highlights the power of polarized neutron data in combination with symmetry analysis.